\documentstyle[pra,epsfig,floats,aps]{revtex}
\input psfig.sty
\tightenlines 
\begin{document}
\title{Statistical Mechanics of Torque Induced Denaturation of DNA}
\author{Simona Cocco $^{1,2,3}$ and R\'emi Monasson $^3$}
\address {$^1$  Dipartimento di  Scienze Biochimiche,
Universit{\`a} di Roma "La Sapienza",
P.le A. Moro, 5 - 00185 Roma, Italy}
\address{$^2$ CNRS-Laboratoire de Physique de l'ENS-Lyon, 
46 All{\'e}e d'Italie, 69364 Lyon Cedex 07, France.}
\address{$^3$ CNRS-Laboratoire de Physique Th\'eorique de l'ENS, 
24 rue Lhomond, 75005 Paris, France.}

\maketitle

\begin{abstract}
A unifying theory of the denaturation transition of DNA, driven by
temperature $T$ or induced by an external mechanical torque $\Gamma$
is presented. Our model  couples  the
hydrogen-bond opening and the untwisting of the helicoidal molecular
structure. We show that denaturation
corresponds to a first-order phase transition from B-DNA to d-DNA phases
and that the coexistence region is naturally parametrized by the
degree of supercoiling $\sigma$.  The denaturation free energy, 
the temperature dependence of the twist angle,  
the phase diagram in the $T,\Gamma $ plane and isotherms in the
$\sigma , \Gamma $ plane are calculated and show a good agreement with
experimental data.
\end{abstract}
\vskip .5cm
PACS Numbers~: 87.14.Gg, 05.20.-y, 64.10.+h
\vskip .5cm

Denaturation of the DNA, due to its essential relevance to
transcription processes has been the object of intensive works in the
last decades.  Experiments on dilute DNA solutions have provided
evidence for the existence of a {\em thermally driven} melting
transition corresponding to the sudden opening of base pairs at a
critical temperature $T_m$ \cite{Saen84}. Later, following the work
of Smith et al.\cite{Smit92}, micromanipulation techniques have been
developed to study single-molecule behaviour under stress conditions
and how structural transitions of DNA can be {\em mechanically}
induced. While most single-molecule experiments have focused on
stretching properties so far, the response of a DNA molecule to an 
external torsional stress has been studied very recently
\cite{Stri99,Chat99}, sheding some new light on denaturation 
\cite{Stri99}. From a biological point of view, torsional stress is 
indeed not unusual in the living cell and may strongly influence DNA
functioning\cite{Stri98,Mark95}.  

For a straight DNA molecule with fixed ends, the degree of supercoiling
$\sigma =(Tw-Tw_0)/Tw_0$ measures the twist $Tw$ ({\em i.e.}  the
number of times the two strands of the DNA double-helix are
intertwined) with respect to its counterpart $Tw_0$ for an
unconstrained linear molecule.  In Strick et al.  experiment
\cite{Stri99}, a $\lambda$ DNA molecule, in 10 mM PB, is attached at
one end to a surface and pulled and rotated by a magnetic bead at the
other end. At stretching forces of $\simeq 0.5$ pN, sufficient to
eliminate plectonems by keeping the molecule straight, a torque induced
transition to a partially denaturated DNA is observed.  Beyond a
critical supercoiling $\sigma_c \simeq -0.015$ and an
associated critical torque $\Gamma_c \simeq -0.05 eV/\hbox{\rm rad}$
the twisted molecule separates into a pure B-DNA phase with 
 $\sigma=\sigma_c$ and denaturated regions with $\sigma= -1$.
Extra turns applied to the molecule increase the relative fraction of
d-DNA with respect to B-DNA.

In this letter, we provide a unifying understanding of both thermally
and mechanically induced denaturation transitions. We show that
denaturation can be described in the framework of first-order phase
transitions with control parameters being the temperature and the
external torque. This is in close analogy to the liquid-gas
transition, where control parameters are the temperature and the
pressure. Our theory gives a natural explanation to the BDNA-dDNA
phases coexistence observed in single molecule experiments
\cite{Stri99}. We give quantitative estimates for
the denaturation free-energy $\Delta G$, the temperature dependence of
 the average twist angle
 $\Delta \langle \theta \rangle /\Delta T$, the critical supercoiling 
$\sigma _ c $ and torque $\Gamma _c$
at room temperature in good agreement with the
experimental data. Furthermore the dependence of the critical torque
as a function of the temperature is predicted.

Our model reproduces the Watson-Crick double helix (B-DNA) as
schematized fig~1. For each base pair ($n=1,\ldots ,N$), we consider a
polar coordinate system in the plane perpendicular to the helical axis
and introduce the radius $r_n$ and the angle $\varphi_n$ of the
base pair \cite{Bar99}. The sugar phosphate backbone is made of rigid
rods, the distance between adjacent bases on the same strand being
fixed to $L =6.9 \AA$. The distance $h_n$ between 
base planes $n-1$ and $n$ is expressed in terms
of the radii $r_{n-1}, r_n$ and the twist angle $\theta _n = \varphi _n
- \varphi _{n-1}$ as
\begin{equation}
\label{h} 
h_n (r _n , r_{n-1} , \theta _n ) =\sqrt{ L^2 - r_n ^2 -
r_{n-1}^2 +2 r_n r_{n-1} \cos \theta _n } \ . 
\end{equation}
The potential
energy associated to a configuration of the degrees of freedom $(r_n ,
\varphi _n )$ is the sum of the following nearest neighbor
interactions.

First, hydrogen bonds inside a given pair $n$ are taken into
account through the short-range Morse potential \cite{Proh95,PB89} 
$V_m (r_n)=D\, \left(e^{-a(r_n-R)}-1\right)^2$ 
with  $R=10\AA$ .
Fixing $a=6.3 \AA ^{-1}$ \cite{Daux95,Zhan97}, the width of the well
amounts to $3a^{-1} \simeq 0.5 {\AA}$ in agreement with the 
order of magnitude of the relative motion of the hydrogen bonded bases 
\cite{Mac87}. A base pair with diameter $r > r_d = R + 6/a$ may be
considered as open. The potential depth $D$, typically of the order of 
$0.1 eV$ \cite{Proh95,Camp98} depends on the base pair type 
(Adenine-Thymine (AT) or Guanine-Citosine (GC)) as well as on the 
ionic strength.

Secondly, the shear force that opposes sliding motion of one base over
another in the B-DNA conformation is accounted for by the stacking
potential \cite{Saen84} $V_s(r_n,r_{n-1})= E\, e^{-b (r_n +
r_{n-1}-2R)}\; (r_n-r_{n-1})^2$.  Due to the decrease of molecular
packing with base pair opening, the shear prefactor is exponentially
attenuated and becomes negligible beyond a distance $\simeq 5 b^{-1} =
10 {\AA}$, which coincides with the diameter of a base pair
\cite{Daux95,Zhan97,Camp98,Cule97}.

Thirdly, an elastic energy $V_b (r _n , r_{n-1} , \theta _n ) = K [
h_n - H ]^2 $ is introduced to describe the vibrations of the molecule
in the B phase. The helicoidal structure arises from $H <
L$: in the rest configuration $r_n=R$ at $T=0$K, $V_b$ is minimum and zero for
the twist angle $\theta_n=2\pi/10$. Choosing $H=3\AA$, we recover
at room temperature $T=298$K the thermal averages $\langle h_n
\rangle \simeq 3.4 \AA$ and $\langle \theta_n \rangle \simeq 2 \pi /
10.4$ \cite{Saen84}. The above
definition of $V_b$ holds as long as the argument of the square root
in (\ref{h}) is positive, that is if $r_n , r_{n-1} , \theta _n$ are
compatible with rigid rods having length $L$. By imposing $V_b =
\infty$ for negative arguments, unphysical values of $r _n , r_{n-1} ,
\theta _n$ are excluded. As the behaviour of a single strand ($r > r_d$)
is uniquely governed by this rigid rod condition, the model does not only
describe vibrations of helicoidal B-DNA but is also appropriate for
the description of the denaturated phase.

As will be discussed later, the elastic constant $K=0.014 eV/\AA^2$ is
determined to give back the torsional modulus $C$ of B-DNA 
estimated to $C=860 \pm 100 {\AA}$ \cite{Croq99,Bouc97} at $T=298$K.
The parameters of the Morse potential $D$ and of the stacking
interaction $E$ we have set to fit the melting
temperature $T_m=350$K of the homogeneous Poly(dGdT)-Poly(dAdC)-DNA
at $20 mM\ Na^+$ \cite{Saen84}, see inset of fig~3.  This melting
temperature coincides with the expected denaturation temperature of a
heterogeneous DNA with a sequence GC/AT ratio equal to unity at $10
mM\ Na^+$ \cite{Saen84}, as the $\lambda$-DNA in the experimental
conditions of \cite{Stri99}. Among all possible pairs of parameters
$(D,E)$ that correctly fit $T_m$, we have selected
the pair $(D=0.16 eV , E=4 eV/\AA^2)$ giving the largest 
prediction for $\Delta G$, see
inset of fig~2, that is in closest agreement with thermodynamical 
estimates of the denaturation free-energy. 

When the molecule is fixed at one end and subject to a torque $\Gamma$
on the other extremity, an external potential $V_{\Gamma}(\theta
_n)=-\Gamma \,\theta _n$ has to be included. A torque $\Gamma
>0$ overtwists the molecule, while $\Gamma <0$ undertwists it.

The configurational partition function at inverse temperature $\beta$
can be calculated using the transfer integral method~:
\begin{equation}
\label{ZG}
Z_{\Gamma}= \int _{-\infty} ^\infty d\varphi _N \; \langle R, \varphi
_N | T^N| R , 0 \rangle
\end{equation}
As in the experimental conditions, the radii of the first and last
base pairs are fixed to $r_1=r_N=R$.  The angle of the fixed extremity
of the molecule is set to $\varphi_1=0$ with no restriction whereas
the last one $\varphi _N$ is not constrained.  The transfer operator
entries read $<r,\varphi|T|r',\varphi'>\equiv X(r,r')\, \exp \{ -\beta
( V_b \,(r,r', \theta ) + V_\Gamma (\theta )) \} \, \chi (\theta )$
with $X(r,r')= \sqrt{rr'} \exp \{-\beta (V_m\,(r)/2+V_m\,(r')/2 +
V_s\,(r,r') ) \}$.  The $\sqrt{rr'}$ factor in $X$ comes from the
integration of the kinetic term; $\chi(\theta) =1$ if $0\leq \theta =
\varphi-\varphi' \leq \pi$ and 0 otherwise to prevent any clockwise
twist of the chain.  At fixed $r,r'$, the angular part of the transfer
matrix $T$ is translationally invariant in the angle variables
$\varphi$, $\varphi '$ and can be diagonalized through a Fourier
transform. Thus, for each Fourier mode $k$ we are left with an
effective transfer matrix on the radius variables
$T_k(r,r')=X(r,r')\,Y_k(r,r')$ with
\begin{equation}
\label{mu}
Y_k(r,r') = 
\int_{0}^{\pi}\,d\theta \,e^{-\beta ( V_b\,(r,r',\theta) +
 V_\Gamma (\theta ) )}e^{-ik\theta} \quad .
\end{equation} 
The only mode contributing to $Z_{\Gamma }$ is $k=0$ once $\varphi _N$
has been integrated out in (\ref{ZG}).  The eigenvalues and
eigenvectors of $T_0$ will be denoted by $\lambda ^{(\Gamma )} _{q}$
and $\psi^{(\Gamma )}_{q}(r)$ respectively with $\lambda^{(\Gamma
)}_{0} \ge \lambda^{(\Gamma )}_{1} \ge \ldots$.  In the $N\to\infty$
limit, the free-energy density $f^{(\Gamma )}$ does not depend on the
boundary conditions on $r_1$ and $r_N$ and is simply given by $f
^{(\Gamma )}= -k_B T \ln \lambda ^{(\Gamma )} _{0}$.


Note that the above result can be straightforwardly extended to the case of a
molecule with a fixed twist number $Tw = N \ell$, e.g. for circular
DNA.  Indeed, the twist density $\ell$ and the torque $\Gamma$ are
thermodynamical conjugated variables and the free-energy at
fixed twist number $\ell$ is the Legendre transform of $f_\Gamma$. 

We have resorted to a Gauss-Legendre quadrature for numerical
integrations over the range $r_{min} = 9.7 {\AA}< r < r_{max}$.
The Morse potential $V_m$ increases exponentially with decreasing $r< R$
and may be considered as infinite for $r<9.7 \AA$ \cite{Zhan97}.
The extrapolation procedure to $r_{max}\to \infty$ depends on the torque
value $\Gamma$ and will be discussed below.
Using Kellog's iterative method \cite{Daux95}, the
eigenvalues $\lambda ^{(\Gamma )} _{q}$ and associated eigenvectors
$\psi ^{(\Gamma )} _{q}(r)$ have been obtained for $q=0,1,2$.
Like a quantum mechanical wave function, $\psi ^{(\Gamma )} _{0} (r)$
gives the probability amplitude of a base pair to be of radius $r$.
Two quantities of interest are:  the
percentage of opened base pairs $P=\int_{r_d}^{\infty} dr\; |\psi
^{(\Gamma )} _{0} (r)|^2 $, the averaged twist angle $\langle \theta 
\rangle = -\partial f_{\Gamma} /\partial \Gamma $.

Results for a freely swiveling molecule at room temperature are as
follows.  $\psi^{(\Gamma=0 )}_{0}$ is entirely confined in the Morse
potential well and describes a closed molecule.  Conversely the
following eigenfunctions $\psi ^{(\Gamma=0 )}_{1}$, $\psi ^{(\Gamma=0
)}_{2}, \ldots$ correspond to an open molecule: they extend up to
$r_{max}$ and vanish for $r<r_d$. They are indeed orthogonal to another
family of excited states that are confined in the Morse potential with
much lower eigenvalues. The shape of the open states are strongly
reminiscent of purely diffusive eigenfunctions, $\psi _q ^{(\Gamma
=0)} (r) \simeq \sin ( q \pi (r - r_d) / ( r_{max} -r_d ))$ leading to
a continuous spectrum in the limit $r_{max} \rightarrow \infty$.

This observation can be understood as follows. For $r,r' > r_d$, the
transfer operator $T_0(r,r')$ is compared fig.~2
to the exact conditional probability $\rho (r,r')$ that the endpoint of a 
backbone rod of length $L$ is located at distance $r'$ from the vertical 
reference axis knowing that its other extremity lies at distance 
$r$ \cite{notar}. For fixed $r$, $T_0$ and $\rho$ both 
diverge in $r'=r\pm L$ and are essentially flat in between. 
The flatness of $T_0$  derive from the expression
of $V_b$: a rigid rod with extremities lying in $r,r'$ may always be 
oriented with some angle  $\theta ^*$ ($\to 0$ at large distances) 
at zero energetic cost 
$V_b (r,r', \theta ^*)=0$.  As a conclusion, our model can reproduce the
purely entropic denaturated phase.

As shown fig~2, at a critical temperature $T_m= 350$K, 
$\lambda ^{(\Gamma = 0)} _{0}$ crosses the second largest
eigenvalue and penetrates, as in a first-order-like transition the
continuous spectrum. For $T> T_m$, the bound state disappears and
$\psi^{(\Gamma=0 )}_{1}$ in fig~2 becomes the eigenmode with
largest eigenvalue \cite{nota4}.  The percentage of opened base pairs
$P$ exhibits an abrupt jump from 0 to 1 at $T_m$, reproducing the UV
absorbance vs. temperature experimental curve for
Poly(dGdT)-Poly(dAdC)-DNA \cite{Saen84}.
The difference $\Delta G$ between the free energy $f_d ^{(\Gamma =0 )}$
of the open state ($q=1$) and the free energy $f_B ^{(\Gamma =0 )}$ 
of the close state ($q=0$) gives the
denaturation free energy at temperature $T$, see fig.~2.  
At $T=298$K, we obtain $\Delta G =0.022 eV$ in good agreement 
with the free energy of the denaturation bubble formation 
$\Delta G \simeq  0.025 eV$ estimated in AT rich regions \cite{Saen84,Stri98}.
The  thermal fluctuations in the B-DNA phase lead to an undertwisting 
$\Delta \langle\theta\rangle /\Delta T\simeq \, -1.4 \, 10^{-4} \hbox{\rm  
rad/K}$ which closely agrees with experimental measures 
$\Delta \langle \theta \rangle /
\Delta T\simeq - 1.7 \, 10^{-4} \hbox{\rm  rad/K}$ \cite{Dep75}.

The presence of an overtwisting (respectively undertwisting) torque
$\Gamma >0$ (resp. $\Gamma <0$) strongly affects $f_B ^{(\Gamma )}$, 
leaving almost unchanged the single strand free-energy $f_d ^{(\Gamma )}$.  
The denaturation transition takes place at $T_m
(\Gamma )$ \cite{nota3}, see the phase diagram 
shown in the inset of fig~3. We expect a critical point at a
high temperature and large positive torque such that
$\psi^{\Gamma}_{1}$ is centered on $R$ \cite{nota3}. 

The supercoiling, induced by a torque at a
given temperature smaller than $T_m (\Gamma =0) = 350$K, is the
relative change of twist with respect to the value at zero torque in
the B-DNA state, $\sigma(\Gamma)= ( \langle \theta \rangle _{\Gamma} -
\langle \theta \rangle _{\Gamma=0} )/\langle \theta \rangle
_{\Gamma=0}$.  In fig~3, we have plotted the isotherms in the $\sigma,
\Gamma$ plane.  Horizontals lines are critical coexistence regions
between the B-DNA phase, on the left of the diagram and the denaturated
phase on the right (with $\sigma=-1$). The left steep line is found 
to define a linear relation between $\Gamma$ and $\sigma$~: $\Gamma 
= {K}_{\theta}\; (\langle \theta \rangle _{\Gamma} - \langle \theta 
\rangle _{\Gamma =0} )$. The slope ${K}_{\theta}$ does not vary
with temperature over the range 298 K$<T<$350 K and is related to the
torsional modulus $C$ of B-DNA through $C={K}_{\theta} \langle h_n \rangle / 
(k_B T)$\cite{Croq99}. The value of $K$ appearing in the elastic
potential $V_b$ and given above was tuned to ensure that 
$C=860\AA$\cite{Croq99,Bouc97}. 
At room temperature, critical coexistence between B-DNA and d-DNA arises
at torque $\Gamma_c=-0.035 eV/\hbox{\rm rad}$ and supercoiling 
$\sigma_c=-0.01$. These theoretical results are in good agreement with
the values $\Gamma_c=-0.05 eV/\hbox{\rm rad}$, $\sigma_c=-0.015$
obtained experimentally\cite{Stri99}.

We plan to combine the present model with existing elasticity theories
of DNA \cite{Smit92,Bouc97} to understand the influence of an external
stretching force on the structural transition studied in this paper.
It would also be interesting to see how the above results are 
modified in presence of a heterogeneous sequence.

{\bf Acknowledgements}~:
The present model is the fruit of a previous collaboration \cite{Bar99}
of one of us (S.C.) with M.~Barbi and M.~Peyrard which we are particularly
grateful to. We also thank B.~Berge, D.~Bensimon, C.~Bouchiat, E.~Bucci, 
A.~Campa, A.~Colosimo, V.~Croquette, A.~Giansanti for useful discussions.

\begin{figure}
\includegraphics[width=150pt,angle=0]{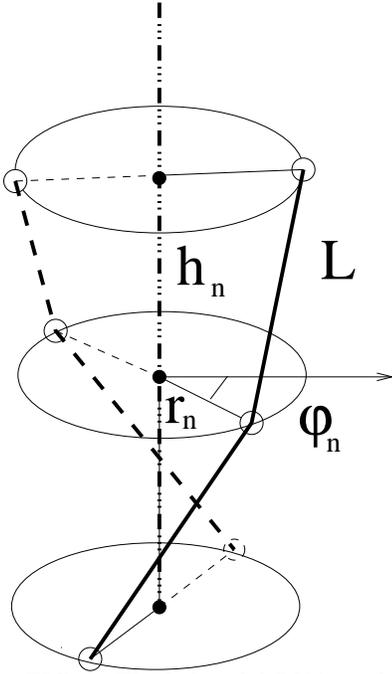}
\caption{The helicoidal DNA model: each base pair
is modelized through its radius $r_n$ and angle $\varphi_n$.
The axial distance $h_n$ between successive base pairs planes varies while 
the backbone length along the strands is fixed to $L$.} 
\protect\label{f1}
\end{figure}

\begin{figure}
\includegraphics[width=200pt,angle=-90]{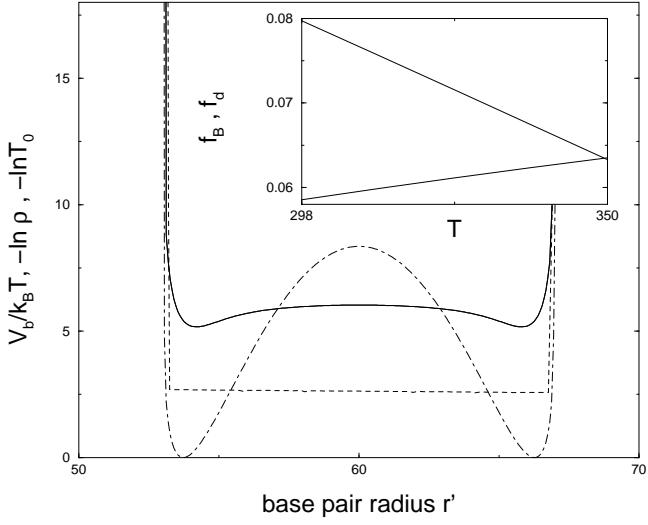}
\caption{With respect to $V_b (r=60,r',\theta =0)$ (dashed-dotted
line, plotted as a function of $r'$ in units of $k_B T$ for $T=298$K),
$- \ln T_0 (r=60,r')$ (full line) 
is flat and comparable to $-\ln \rho (r=60, r')$
(dashed line) up to an irrelevant additive constant.
Similar curves are obtained for any $r>r_d$.
Inset: Free-energies $f_B ^{(\Gamma =0)}$ (lower curve) 
and $f_d ^{(\Gamma =0)}$ (upper curve) of B- and d-DNA.
$\Delta G$ equals $0.022 eV$ at $T=298$K 
and vanishes at $T_m=350$K.}
\protect\label{f2}
\end{figure}

\begin{figure}
\includegraphics[width=200pt,angle=-90]{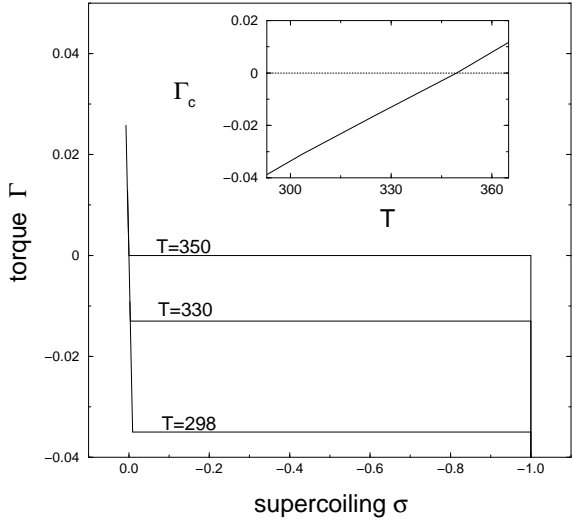}
\caption{Isotherms in the  $\sigma, \Gamma$ plane. For $T=298$K,
the critical  supercoiling and  torque are $\sigma_c=-0.01 $,
$\Gamma_c=-0.035 eV/rad$ respectively. Inset~: phase diagram in the
$T, \Gamma$ plane. At $T=298$K, $\Gamma_c=-0.035 eV/rad$ while in the
absence of torque (dotted line), $T_m=350$K.}
\protect\label{f3}
\end{figure}


\begin{thebibliography}{99}
\bibitem{nota3}
$\psi _1 ^{(\Gamma )}$ is centered around $r< r_{max} /2 $ (resp.  $r>
r_{max} /2 $) for $\Gamma >0$ (respectively $\Gamma <0$).  
For positive or zero torque, the melting temperature $T_m (r_{max})$ 
remains unchanged as soon as $r_{max} > 150-200
{\AA}$. When $\Gamma <0$, $T_m (r_{max}) \simeq T_m + c/ r_{max
}$. $c$ increases with the torque strength and is related to
the pressure exerted by the system walls.
\end{thebibliography}
\end{document}